\documentclass{article}

\usepackage{arxiv}

\usepackage{listings}

\usepackage{siunitx}
\usepackage{verbatim}
\usepackage{hyperref}
\usepackage{listings}
\usepackage{xcolor}
\usepackage{pgfplots}
\usepackage{cite}
\usepackage{amsmath,amssymb,amsfonts}
\usepackage{graphicx}
\usepackage{textcomp}
\usepackage{xcolor}
\def\BibTeX{{\rm B\kern-.05em{\sc i\kern-.025em b}\kern-.08em
    T\kern-.1667em\lower.7ex\hbox{E}\kern-.125emX}}

\usepackage{algorithm}
\usepackage[noend]{algpseudocode}

\usepackage{xspace}

\newboolean{showcomments}
\setboolean{showcomments}{true}
\ifthenelse{\boolean{showcomments}}
{ \newcommand{\mynote}[3]{
    \fbox{\bfseries\sffamily\scriptsize#1}
    {\footnotesize$\blacktriangleright$\textsf{\emph{\color{#3}{#2}}}$\blacktriangleleft$}}}
{ \newcommand{\mynote}[3]{}}

\newcommand{\LONG}[1]{}
\newcommand{\SHORT}[1]{#1}

\newcommand{\new}[1]{\mynote{new}{#1}{brown}}

\makeatletter
\def\BState{\State\hskip-\ALG@thistlm}
\makeatother

\definecolor{codegreen}{rgb}{0,0.6,0}
\definecolor{codegray}{rgb}{0.5,0.5,0.5}
\definecolor{codepurple}{rgb}{0.58,0,0.82}
\definecolor{backcolour}{rgb}{0.95,0.95,0.92}

\lstdefinestyle{mystyle}{
    backgroundcolor=\color{backcolour},   
    commentstyle=\color{codegreen},
    keywordstyle=\color{magenta},
    numberstyle=\tiny\color{codegray},
    stringstyle=\color{codepurple},
    basicstyle=\ttfamily\footnotesize,
    breakatwhitespace=false,         
    breaklines=true,                 
    captionpos=b,                    
    keepspaces=true,                 
    numbers=left,                    
    numbersep=5pt,                  
    showspaces=false,                
    showstringspaces=false,
    showtabs=false,                  
    tabsize=2
}

\newcommand{\reliablebroadcast}{RBcast\xspace}
\newcommand{\action}[1]{{\small\texttt{#1}}}
\newcommand{\event}[1]{{\textit{#1}}}
\newcommand{\type}[1]{{\textit{#1}}}
\newcommand{\state}[1]{{\textit{#1}}}

\title{Learning to generate Reliable Broadcast Algorithms
}

\date{July 29, 2022}	

\author{ 
	Diogo Vaz, David R. Matos, Miguel L. Pardal, Miguel Correia  \\
	INESC-ID, Instituto Superior Técnico, Lisboa -- Portugal \\
	\{diogo.vaz, david.r.matos, miguel.pardal. miguel.p.correia\}\texttt{@tecnico.ulisboa.pt} \\
}

\renewcommand{\shorttitle}{Learning to generate Reliable Broadcast Algorithms}

\begin{document}
\maketitle


\begin{abstract}
Modern distributed systems are supported by fault-tolerant algorithms, like \emph{Reliable Broadcast} and \emph{Consensus}, that assure the correct operation of the system even when some of the nodes of the system fail.
However, the development of distributed algorithms is a manual and complex process, resulting in scientific papers that usually present a single algorithm or variations of existing ones. 
To automate the process of developing such algorithms, this work presents an intelligent agent that uses \emph{Reinforcement Learning} to generate correct and efficient fault-tolerant distributed algorithms.
We show that our approach is able to generate correct  fault-tolerant \emph{Reliable Broadcast} algorithms with the same performance of others available in the literature, in only $12,000$ learning episodes.
\end{abstract}




\section{Introduction}
\label{section:introduction}

Distributed systems are made of multiple components interconnected by communication networks.
By leveraging  distribution, such systems can provide high {availability} and {scalability}. Examples of modern and widely used distributed systems are 
cloud applications\cite{armbrust2010view} 
and  blockchains\cite{willChangeTheWorld}. 
During normal operation, some of these components may {fail}, e.g., due to power loss, software bugs, or malicious attacks. These faults can compromise the normal functioning of the entire system. Therefore, it is necessary to provide fault tolerance properties so that the distributed system can 
maintain its normal execution, even in the presence of faults. 
For that purpose, it is important to design and implement \textit{fault-tolerant distributed algorithms}\cite{cachin2011introduction,lynch1996distributed}.

Fault-tolerant distributed algorithms have been widely studied and developed over the years \cite{ben1983another,bonomi2021practical,bracha1984asynchronous,castro1999practical,correia2006consensus,correia2010asynchronous,imbs2015simple}, 
exploring different aspects such as 
the problem to study
\cite{imbs2015simple,castro1999practical}, 
the failures mode\cite{hadzilacos1994modular,correia2010asynchronous} 
or the system architecture \cite{correia2006consensus,bonomi2021practical}. 
However, the process of 
studying and designing a fault-tolerant algorithm is a \textit{manual} and \textit{complex} endeavor\cite{hadzilacos1994modular}. This is especially true in the presence of malicious actors, where the algorithms are very subtle, complicated and slight changes often require a complete redesign of the algorithm. Moreover, current research on fault-tolerant algorithms do not focus only on \emph{correctness}: \emph{efficiency} is also a very important concern in these algorithms, which increases the complexity of the overall algorithm development process. 

Typically, the journey to develop a fault-tolerant algorithm starts with assumptions about the environment or system model, e.g., if the system is {synchronous} or {asynchronous} \LONG{(i.e., if there are assumptions about time or not)} and the failures that can occur.
Researchers then have to define the problem they want to solve, e.g., Reliable Broadcast, shortened in this paper as \reliablebroadcast\LONG{ or Consensus}. 
Next, researchers think about the strategy to design the distributed algorithm. In this stage, besides the difficulty of creating the algorithm, researchers may be biased by previous related papers and algorithms. After a trial-and-error process, the distributed algorithm is generated. This also involves a validation process to assess whether the generated algorithm achieves the goal and solves the problem correctly. This can be done by writing a proof and/or by doing verification using a model checker or a theorem prover.

In this work, our aim is to automate the process of generating fault-tolerant distributed algorithms by proposing an \emph{intelligent agent}\cite{norvig2002modern} capable of generating and validating 
algorithms for a specific distributed problem.\footnote{We take the term \emph{intelligent agent} from the Artificial Intelligence area, where Reinforcement Learning appeared. If the term ``intelligent'' makes sense is, of course, debatable.} More precisely, we aim to create an agent that can generate correct and efficient algorithms based on the \emph{inputs} given by researchers, i.e., assumptions about the environment, failures, and characteristics of the algorithm. 
Our solution aims to be \emph{flexible} so that researchers can change the specification with the objective of studying different problems that, consequently, can lead to different algorithms.
This is a novel research area, so in this paper we start with 
a single problem: \reliablebroadcast, but our 
work aims to be applicable to other distributed problems.

Our approach follows 
a process with two phases: 
\emph{generation}
, for obtaining candidate algorithms, and 
\emph{validation}
, for assessing their correctness.
The \emph{generation} process is related to the problem of code/program generation but with significant differences, 
as distributed algorithms are executed in parallel in several nodes and are subject to faults, whereas local programs are not. 
Research on automatic code generation explored static techniques such as design patterns~\cite{budinsky1996automatic}, UML~\cite{moreira2010automatic}, and reverse engineering~\cite{orvalho2020squares} but lately, supervised machine learning techniques have started to be used~\cite{allamanis2018survey,acsirouglu2019automatic,zhang2019analysis} to improve the process.
In this work, we use an \emph{unsupervised machine learning} approach called Reinforcement Learning~\cite{kaelbling1996reinforcement,sutton2018reinforcement,wiering2012reinforcement}. 
To the best of our knowledge, this work is the first to present such an approach on the automatic code generation field.

In terms of \emph{validation}, the search for a validation process for distributed algorithms has been pursued for years~\cite{gmeiner2014tutorial,goel2021towards,lamport1994specifying,rahli2018velisarios}. We identified four possible languages and frameworks to be used: the {TLA+} language and tools\cite{lamport1994specifying}, the {Spin} framework with the {PROMELA} language\cite{john2013towards}, the {ByMC} framework\cite{konnov2018bymc}, and  {IC3PO}\cite{goel2021ic3po}. We opted to use Spin/PROMELA, as explained later.

The main contributions of the paper are: 
(1) a new approach for 
generating correct and efficient distributed fault-tolerant \reliablebroadcast algorithms using machine learning, instead of manual development by human beings, 
(2) 
an intelligent agent 
to generate such distributed algorithms and
(3) an experimental evaluation of the approach and the agent, showing a correct generation of \reliablebroadcast algorithms\LONG{and a new $\lfloor(N-1)/3\rfloor$ Byzantine-tolerant algorithm}.


\section{Reliable Broadcast}
This section presents \reliablebroadcast, the distributed problem for which we want to find algorithms to solve.


\subsection{System Model}
\label{section:system_model}

The system model considered for the \reliablebroadcast algorithms we want to generate is inspired on the modular approach to fault-tolerant problems by Hadzilacos and Toueg
~\cite{hadzilacos1994modular}. 
The system is composed of a static group of \emph{N} processes, i.e., there are no joins or leaves during execution. We assume a \emph{fully connected point-to-point} network, i.e., that all processes are connected with each other through 
\emph{links} and communicate by passing messages. 
We also assume that the system is \emph{asynchronous}, i.e., that the communication delays are neither upper bounded nor respect a global stabilization time. 

\LONG{  

Figure \ref{fig:system_architecture} exemplifies a system composed of three processes \emph{A}, \emph{B}, and \emph{C}. All processes interact with two different buffers: an \emph{input message buffer} (imb) that receives messages, 
and an \emph{output message buffer} (omb) that contains messages to be sent: 
each process omb is connected with every other imb of the system, through a {link}. Next we analyze 
each system component in depth.

\begin{figure}
    \centering
    \includegraphics[width=0.8\columnwidth]{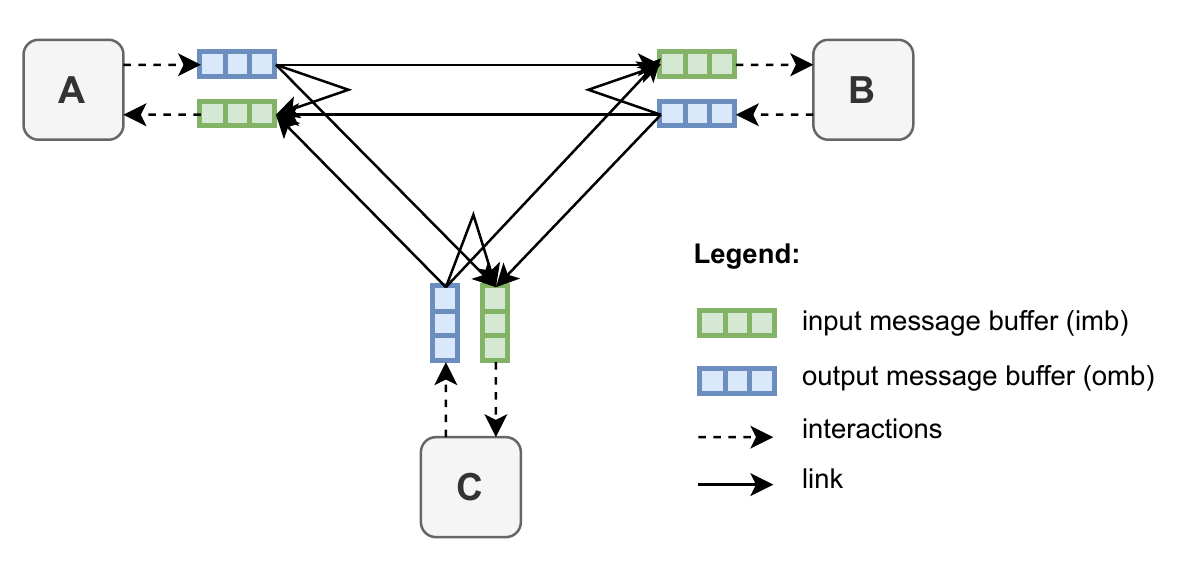}
    \vspace{-0.2cm} 
    \caption{Example of a system with three processes: A, B and C.}
    \label{fig:system_architecture}
\end{figure}

} 


\LONG{\subsection{Processes}}
\label{section:process}

A \textit{process} is the actor of the distributed system, that executes a set of specific ordered \textit{actions}, designated an \emph{algorithm}.
All processes of the system execute the same algorithm.


\LONG{\subsection{Links}}
\label{section:links}

The communication links allow processes to exchange \emph{messages}. 
The links have the task of transporting the message from the output buffer of the sender to the input buffer of the receiver. We assume that the links are \emph{reliable}, \emph{authenticated}, and \emph{provide integrity on the messages}, meaning that there are no corrupted, lost or duplicated messages. However, messages may arrive out of order. 


\LONG{\subsection{Messages}}
\label{section:messages}

Processes use \emph{messages} to share data between them. Typically, a message contains data such as the message content (used by the logic of the system) and an identifier that can contain \emph{protocol type}, \emph{sender}, and \emph{sequence number}. 


\LONG{\subsection{Failure Modes}}
\label{section:failure_modes}

The processes of the system can be correct or faulty. 
We consider \emph{three failure modes}: No-Failure, Crash-Failure, and Byzantine-Failure. 
In the simplest, \emph{No-Failure}, we assume there are no failures. 
%
%
In the \emph{Crash-Failure mode} \cite{bazzi2001simplifying,dwork1990knowledge}, processes may stop operating and never recover. 
Assuming a system with $N \in \mathbb{N}$ processes, the maximum number of faulty processes $F \in \mathbb{N}$ due to a crash failure that can be tolerated in the system is $F=\lfloor(N-1)/2\rfloor$ \cite{bracha1985asynchronous}.
In the \emph{Byzantine-Failure mode} \cite{lamport1982byzantine} faulty processes may have arbitrary behavior, e.g., they may execute other actions not defined by the algorithm or even not execute any action at all. Unlike a crash failure, when a process suffers a Byzantine failure (or, ``is Byzantine''), it can continue to work. 
Assuming a system with $N$ processes, the maximum number of faulty processes $F$ due to a Byzantine failure that can be tolerated in the system is $F=\lfloor(N-1)/3\rfloor$ \cite{bracha1985asynchronous,correia2006consensus}.


\subsection{Problem Definition}
\label{section:reliable_broadcast}

A \reliablebroadcast algorithm ensures, essentially, that every message broadcast by a \emph{correct} process (an \event{RB-Broadcast} event) is eventually delivered by all \emph{correct} processes. The protocol is defined by the following properties\cite{cachin2001secure,hadzilacos1994modular}:

\begin{itemize}
    \item \emph{RB-Agreement}: if a correct process delivers a message \emph{m}, then all correct processes will eventually deliver the same message \emph{m};
    \item \emph{RB-Validity}: if a correct process broadcasts a message \emph{m}, then it will eventually deliver that message \emph{m};
    \item \emph{RB-Integrity}: for any message \emph{m}, every correct process delivers \emph{m} at most once and only if \emph{m} was previously broadcast by some correct process. 
\end{itemize}

The term \emph{correct} 
refers to a process that follows the algorithm. Otherwise, we call it incorrect or faulty. 

\LONG{ 
Sometimes a system requires stronger properties than these, e.g., properties about the order of message delivery. To provide these extra properties, variants of the \reliablebroadcast problem were specified, such as: \emph{FIFO Broadcast}, where messages broadcast by the same process are delivered in the order they were broadcast;  \emph{Atomic Broadcast}, that assures all processes deliver the same messages in the same order \cite{hadzilacos1994modular,johnen2021fifo,correia2006consensus}. Nevertheless, in this work we use only the default \reliablebroadcast to illustrate our approach to automated algorithm generation.
}


\subsection{Algorithms}
\reliablebroadcast algorithms can differ. 
In this section, we present the definitions of the algorithm adopted in this paper, namely, the structure, messages, types, conditions and efficiency.

\subsubsection{Structure}
\label{section:algorithm_structure}

\reliablebroadcast has been much studied over the years \cite{ben1983another,bracha1984asynchronous,correia2006consensus,imbs2015simple,imbs2016trading,bonomi2021practical}. The papers present \reliablebroadcast algorithms with different structures. For example, Bracha and Toueg presented algorithms organized in execution steps \cite{bracha1984asynchronous,bracha1985asynchronous}, whereas more recent work favors a structure in terms of event handling routines \cite{imbs2015simple,imbs2016trading}. We follow the latter, event-oriented, structure. 

We assume that the structure of the algorithm is composed of two events: the \event{RB-Broadcast} event, triggered only once by the process that starts the execution of the algorithm; and a \event{receive} event, triggered every time a process receives a message. Each event can contain a set of \emph{actions}, i.e., instructions that are executed when the event is triggered. 
The execution of the algorithm ends when there are no more messages to be received.






\subsubsection{Messages and Types}
\label{section:messages_and_type}

As previously discussed in Section~\ref{section:messages}, messages typically contain multiple elements, e.g., content, sender, and protocol type. However, to solve the \reliablebroadcast problem, it is necessary to add a new element, called \emph{type}, to differentiate between the different communication steps of the algorithm. In this work, we follow the definition of previous \reliablebroadcast algorithms~\cite{bracha1984asynchronous,imbs2016trading}, by presenting the message in the format \emph{\textless t,m\textgreater}, where \emph{t} symbolizes the type of the message, and \emph{m} the message itself. Previous works define the types as words like \emph{echo} or \emph{init} \cite{bracha1984asynchronous,imbs2015simple}, but in this work, we designate types 
\type{type0}, \type{type1}, \type{type2}, etc. The reason for this option is that the types are generated automatically, not by a human. But afterwards, we can translate the type 
\type{type0} into \type{init} and the type \type{type1} into \type{echo}.


\subsubsection{Conditions}
\label{section:conditions}


Conditions are statements used to evaluate when a specific action can be executed and are 
associated with the \emph{if} clause. 
On \reliablebroadcast, a condition is 
defined by two properties: the \emph{message \textless t,m\textgreater} and the \emph{threshold}, which is the number of messages needed to satisfy the condition. For example, the condition 
\emph{if received (\textless type0,m\textgreater) from $F+1$ distinct processes} 
means that the action can only be executed if the process has received a message \emph{m} of \emph{type0} from at least $F+1$ processes (the threshold is $F+1$). 


\subsubsection{Efficiency} 
\label{section:algorithm_evaluation}


To analyze the efficiency of an algorithm, we adopt a model, based on previous works\cite{cachin2011introduction,raynal2018fault}, where the efficiency is related to three properties: 
(1) the number of messages sent by the algorithm;
(2) the number of 
communication steps (which in the case of \reliablebroadcast is the number of types, as there are no loops) and; 
(3) the number of messages that have to be received for the algorithm to stop. 
All these metrics indirectly express the cost of computational power, storage and network to execute the algorithm. As the algorithm sends more messages or contains more communication steps, then the process will need more storage for messages, spend more network resources, and take more time to execute the algorithm.




\section{Learning the \reliablebroadcast Algorithm}
\label{section:learning_distributed_algorithms}


\begin{figure}
    \centering
    \includegraphics[width=0.8\columnwidth]{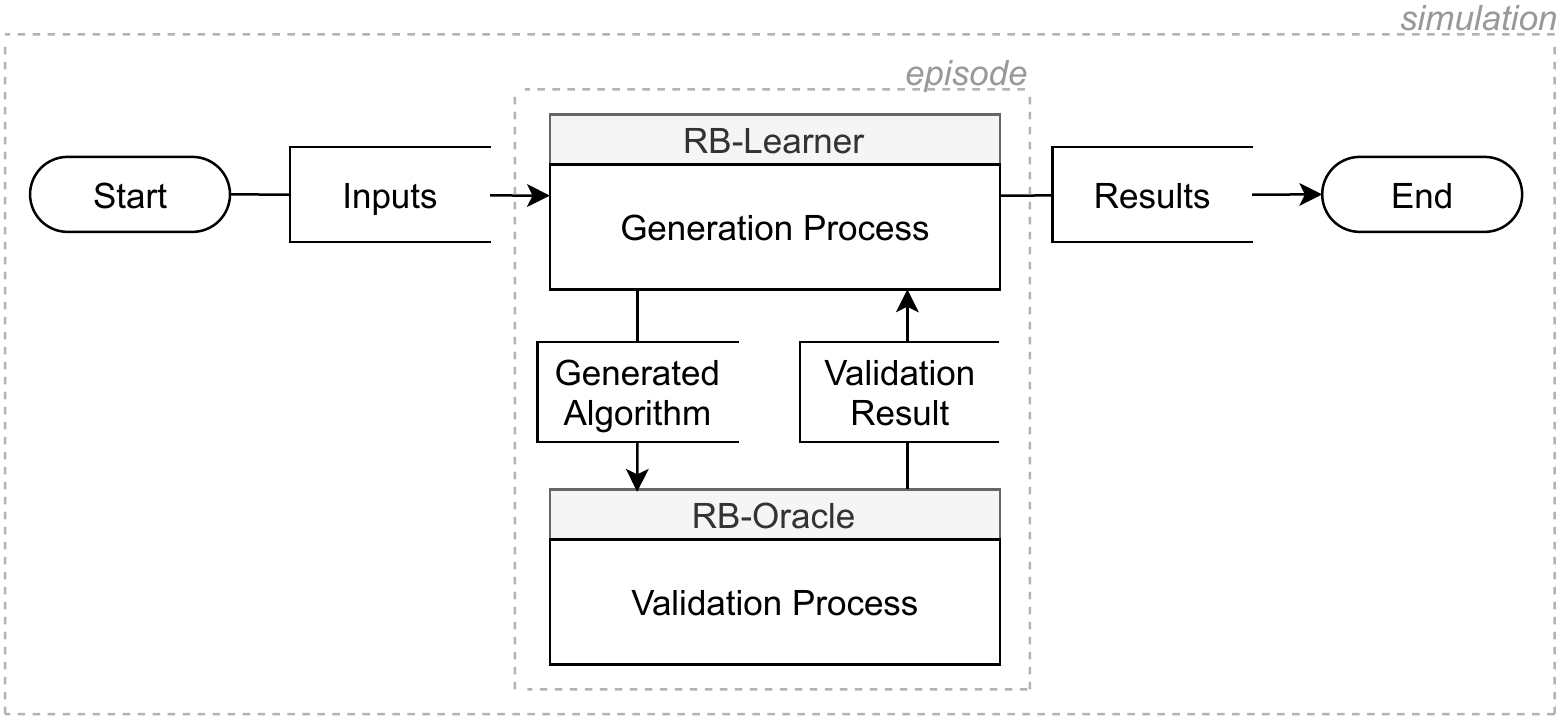}
    \caption{Process/dataflow of generating one algorithm. One simulation involves many episodes.}
    \label{fig:solution_architecture}
\end{figure}



\begin{table}[h]
\caption[Simulation inputs.]{Simulation inputs.} 
\centering 
\resizebox{14cm}{!}
{
\begin{tabular}{ l l | l l}
\hline 
\textbf{Generation process inputs} & \textbf{Possible Values} & \textbf{Validation process inputs} & \textbf{Possible Values} \\ [0.5ex] 
\hline 
Number of simulations & \{1,2,3,...\} & Failure modes to model & \{No-Failure,Crash-Failure, Byzantine-Failure\} \\
Number of episodes & \{1,2,3,...\} & Number $N$ of nodes to model & \{3,4,5,...\} \\
Rewards & \{...,-2,-1,0\} & Fault tolerance ratios & \{0,(N-1)/2, (N-1)/3\} \\
& & & for No-/Crash-/Byzantine-Failure, respectively \\
Heuristics to be applied & (cf.\ Sec.\ \ref{section:generation_heuristics}) & Properties to be validated & \{RB-Validity, RB-Agreement, RB-Integrity\} \\
& & Event handlers where faults can occur & \{RB-Broadcast, receive\} \\
\hline 
\end{tabular}
}
\label{table:inputs} 
\end{table}


Machine learning techniques can be divided into two broad classes: \emph{supervised machine learning} -- to learn from labeled data -- and \emph{unsupervised machine learning} -- to learn without labeled data. 
In this task of learning how to solve a distributed problem, we decided to emulate the \emph{trial and error} strategy used by a human researcher to solve this problem. Therefore, we decided to use an unsupervised machine learning technique called \emph{Reinforcement Learning}
~\cite{kaelbling1996reinforcement,sutton2018reinforcement,wiering2012reinforcement}.

Reinforcement Learning is based on the idea of an \emph{agent} choosing \emph{actions} in specific \emph{states}. The states are the observations that the agent receives from the \emph{environment} where it acts.
The way in which the agent acts is defined by the \emph{policy}, in this case, a map from perceived states to actions to be taken in those states. Then, by choosing an action in a state, the agent will receive a \emph{reward}, reflecting its choice. With time, the agent will start learning the \emph{value} of each state, i.e., the total amount of reward the agent can expect to accumulate over the future, starting from that state. While rewards are \emph{short-term} indicators, the values reflect the \emph{long-term} desirability of that state, taking into account the states that are likely to follow and the rewards associated with them. We consider a \emph{model-free} Reinforcement Learning problem, meaning that the agent does not create a representation of the behavior of the environment, 
unlike model-based approaches. A model-based agent would needed to have a model of the dynamics of the environment, allowing to predict state transitions and rewards, e.g., giving the current algorithm, the agent could plan future actions and their rewards.

Our solution considers an \emph{agent} that has the goal of generating correct and efficient \reliablebroadcast algorithms, i.e., algorithms that satisfy the \reliablebroadcast properties (cf.\ Section \ref{section:reliable_broadcast}) and minimizes the efficiency metrics (cf.\ Section \ref{section:algorithm_evaluation}). The solution is composed of a main agent, the \emph{RB-Learner}, that collaborates with an auxiliary
agent, the \emph{RB-Oracle}, as represented in Figure~\ref{fig:solution_architecture}. 
The entire execution of the solution for one algorithm is designated a \emph{simulation}.
The process starts with the definition of the \emph{inputs} of the learning process. Table \ref{table:inputs} summarizes these inputs. 
The \emph{generation process inputs} include the number of simulations and episodes to run and the rewards and heuristics to be used, the last two containing domain knowledge about the problem (see Sections \ref{section:runtime_rewards} and \ref{section:generation_heuristics}). The \emph{validation process inputs} include the specifications to validate the algorithm, such as the failure modes and their tolerance ratios, the number of nodes to model, the properties and the event handlers to be validated. Then, the execution starts with the RB-Learner 
\emph{generation process}. After that, the RB-Learner gives the generated algorithm to the RB-Oracle, 
executing the \emph{validation process}, i.e., to assess whether the generated algorithm actually solves the problem. Then, the RB-Oracle returns the validation result to the RB-Learner, which it uses as part of the learning process. 
These iterations RB-Learner $\rightarrow$ RB-Oracle $\rightarrow$ RB-Learner $\rightarrow \dots$ are repeated many times. Each iteration is designated an \emph{episode}.
In the end, the RB-Learner outputs the results of the execution: the most efficient algorithm generated. 

With this technique, the idea is that the agent will be able to learn/generate a correct and efficient algorithm by generating multiple algorithms -- either correct or incorrect -- without the need of prior knowledge of the state of the art in \reliablebroadcast. 


\section{RB-Learner}
\label{section:rb_learner}

The RB-Learner agent uses Reinforcement Learning to learn 
not only 
an algorithm that solves the problem, but also 
an algorithm that is efficient. Next, we explain the elements behind the learning process of the RB-Learner. 



\begin{table*}[t]

\caption[Actions available to the RB-Learner.]{Actions available to the RB-Learner.} 
\centering 
\resizebox{13cm}{!}
{
\begin{tabular}{l l | l l}
\hline 
\textbf{Logic} & \textbf{Message type}& \textbf{Condition threshold} & \textbf{Message type} \\ [0.5ex] 
\hline 
\action{SEND to all}(\textless t,m\textgreater) &  \type{type0}, \type{type1} & $0$, $1$, $F+1$, $(N+F)/2$, $N-F$ & \type{type0}, \type{type1} \\ 
\action{SEND to neighbours}(\textless t,m\textgreater)  &  \type{type0}, \type{type1} & $0$, $1$, $F+1$, $(N+F)/2$, $N-F$ & \type{type0}, \type{type1}  \\ 
\action{SEND to myself}(\textless t,m\textgreater)&  \type{type0}, \type{type1} & $0$, $1$, $F+1$, $(N+F)/2$, $N-F$ & \type{type0}, \type{type1}\\ 
\action{DELIVER}(m) &  - & $0$, $1$, $F+1$, $(N+F)/2$, $N-F$ & \type{type0}, \type{type1}\\ 
\action{STOP} &  - & $0$ & \type{type0} \\ 
\hline 

\end{tabular}
}
\label{table:actions_available} 
\end{table*}


\subsection{Actions}
\label{section:actions}

An algorithm is composed of a set of event handling routines -- two in the case of \reliablebroadcast, each of which contains a sequence of actions.  
The RB-Learner selects an \emph{action} from the set of possible actions to add to one of the routines. Each action has two components: the \emph{logic} -- the part of the action that is executed -- and the \emph{condition} -- a statement that must be true in order for the logic to be executed. For example, in action 

\begin{center}
{\textcolor{red}{\action{SEND to all}(\textless type1,m\textgreater)} \textcolor{blue}{if received (\textless type0,m\textgreater) from 1 distinct process}}
\end{center}

\noindent
the logic component is the left-hand part (in \textcolor{red}{red})  and the condition is the right-hand part (in \textcolor{blue}{blue}, starting with the word ``if''). 
Next, we present the logic and condition components 
that we assume in the paper.

\subsubsection{Logic}

We selected the following logic components taken from previous works that solve the \reliablebroadcast problem\cite{bracha1984asynchronous,hadzilacos1994modular,imbs2015simple}: 
\begin{itemize}
    \item {\action{SEND to all}(\textless t,m\textgreater)}: sends the message \emph{\textless t,m\textgreater} to all processes of the system (including itself);
    \item {\action{SEND to neighbors}(\textless t,m\textgreater)}: sends the message \emph{\textless t,m\textgreater} to all the processes of the system (excluding itself); 
    \item {\action{\action{SEND to myself}}(\textless t,m\textgreater)}: sends the message \emph{\textless t,m\textgreater} only to itself; 
    \item {\action{DELIVER}(m)}: delivers the message \emph{m};
    \item {\action{STOP}}: end the execution of the event handler. 
\end{itemize}

In addition to these, we could think of a generic component to send a message to $N-X$ processes, for any $X>0$ and $X \in \mathbb{N}$. However, this generality is found only in probabilistic algorithms \cite{cachin2011introduction,eugster2003lightweight}, which are outside the scope of this work.


\subsubsection{Conditions}
\label{section:actions_conditions}

In this work, each action is associated with a specific \emph{condition} that defines when the action is executed, and each condition is associated with a specific \emph{threshold}. 
From a range of works analyzed \cite{bracha1984asynchronous,hadzilacos1994modular,imbs2015simple,raynal2018fault}, we selected five thresholds: waiting for \emph{0} (always true, a tautology), $1$, $F+1$, $(N+F)/2$, and $N-F$ messages from different processes. 
We assume that two conditions are equal if they wait for the same message type and have the same threshold. 


\subsubsection{Message type}
\label{section:actions_message_types}

As explained above, messages contain a type: \type{type0}, \type{type1}, and so forth.
One parameter of the problem of generating an algorithm is how many types of messages it uses. Clearly, the minimum number of types found in algorithms in the literature corresponds to the maximum number of types needed to solve the problem. 
For \reliablebroadcast we found that the number is of two types \cite{imbs2015simple}.
Both \action{SEND} actions and conditions are influenced by a certain type (except conditions with $threshold=0$ that wait for no messages).



Table \ref{table:actions_available} presents all actions available to our agent, 
resulting in a total of $\mathit{T} =64$ possible actions -- the total number of possible combinations using the components of the table. All actions are associated with all possible conditions except \action{STOP} that does not depend on messages being received (or, equivalently, it depends only on the tautology).

We also assume that each action contains an implicit condition that forbids the action to execute more than once for the same message. This means that if a process delivers a message with content $m=1$, it will never deliver the same message again. The same applies to \action{SEND} actions. This inner condition is introduced by previous articles that present \reliablebroadcast algorithms: for example, in \cite{imbs2015simple} we have the conditions \emph{not yet broadcast} or \emph{not yet RB\_delivered} and in \cite{hadzilacos1994modular} we have the condition \emph{if p has not previously executed deliver(R,m)}.


\subsection{States}
\label{section:states}

A typical Reinforcement Learning agent interacts with an external \emph{environment}. In our case, the environment is not external, but internal memory. 
This memory stores the actions already selected by the agent to form the algorithm. Specifically, a state is the sequence of actions selected by the agent up to that moment. By following this representation, the agent will be able to learn to select the best actions based on the actions that the algorithm already contains. Each state follows the algorithm structure defined in Section \ref{section:algorithm_structure}, being composed by two event handlers and expressed as \state{State([])}. We assume that \state{State} A and \state{State} B are equal 
if \emph{both contain the same actions, in the same number, in the same event handlers}. The order of the messages inside an event handler is not relevant.


\subsection{Rewards}
\label{section:runtime_rewards}

Rewards are used by the agent to learn which actions are suitable or not in each state -- a technique called \emph{reward shaping}. In this work, the rewards are related with the \emph{efficiency} (cf.\ Section \ref{section:algorithm_evaluation}) and \emph{correctness} (cf.\ Section \ref{section:reliable_broadcast}) of the algorithm: the most efficient correct algorithm will generate the best reward. Rewards are defined as 
part of the input, but it 
is important to mention that the agent generates the most efficient algorithm not because of the absolute reward values that we have defined, but because of the relative values between them (the absolute rewards are the result of testing multiple possibilities).

The RB-Agent receives a reward in two moments: (1) every time the agent selects an action -- \emph{runtime reward} -- and (2) when the agent receives the validation result from the RB-Oracle -- \emph{bonus reward}.



\subsubsection{Runtime rewards}

Runtime rewards are related with the \emph{efficiency} of the produced algorithm, i.e., more efficient algorithms will generate the best rewards. Table \ref{table:actions_rewards} summarizes the values we empirically established for calculating these rewards. 

The \action{SEND} actions and the \action{DELIVER} action have a negative reward, as processes need to spend time and resources to execute these actions, so there is a cost involved. For the \action{SEND} actions, the \action{SEND to myself} action has a better reward than the \action{SEND to neighbours} and the \action{SEND to all}  actions. This happens due to the number of sent messages metric: \action{SEND to myself} only sends $1$ message, whereas the others involve sending $N-1$ and $N$ messages, respectively. 
%
The threshold of the conditions 
also influences the reward. 
Table \ref{table:actions_rewards} shows the rewards associated with each selected threshold (c.f Section \ref{section:actions_conditions}), 
following the idea that $0<1 \leq  F+1 \leq (N+F)/2 \leq N-F$, where N is the total number of processes and F the maximum number of faulty processes. 

Beyond what is presented in Table \ref{table:actions_rewards}, the addition of a new message type to the algorithm also involves a (negative) reward. Specifically, the reward is added when a \action{SEND} action introduces a new type. Each new type has an increasing negative reward: \type{type0} is associated with the reward $0$, \type{type1} is associated with reward $-1$, etc. The objective is for the agent to add the minimum number of new types to the algorithm, as each new type involves more communication.


The last aspect that influences the reward obtained by the agent is the event handler where the action is selected. 
We defined that each action selected for the \event{RB-Broadcast} event handler has an additional reward of $0$, while the actions selected for the \event{receive} event handler have an additional reward of $-1$. 
These rewards favor the addition of actions to the \event{RB-Broadcast} event handler instead of to the \event{receive} event handler; this bias is needed because \event{RB-Broadcast} is executed only once per execution of the algorithm, whereas \event{receive} is executed $N$ times (one per process), meaning that an action on the \event{receive} event handler will have a greater impact on the efficiency of the algorithm when compared to an action on the \event{RB-Broadcast} event handler, e.g, a \action{SEND to myself} action will have a cost of 1 message if executed on the \event{RB-Broadcast} event handler, but a cost of $N*1$ if executed on the \event{receive} event handler.

To summarize, consider the example where the agent chooses the action:

\begin{center}
{\action{\textcolor{red}{SEND to all}}(\textless \textcolor{blue}{type0},m\textgreater) if received (\textless type0,m\textgreater) \\from \textcolor{brown}{$N-F$} distinct parties}
\end{center}

The reward for this action will be \textcolor{red}{$-3$} (the \action{SEND to all} logic) 
+ \textcolor{blue}{$0$} (\type{type0} sent)
\textcolor{brown}{$-4$} (the $N-F$ threshold)
= $-7$. Then, if the action is selected for the \event{RB-Broadcast} event handler, it will receive an additional reward of $0$ (still a total of $-7$), while if selected for the \event{receive} event handler, it will receive an additional reward of $-1$ (total of $-8$). 


\subsubsection{Bonus rewards}
\label{section:bonus}


After the algorithm is validated by the RB-Oracle, the RB-Learner receives the validation result from the RB-Oracle. 
The RB-Learner will use that result -- correct or incorrect -- to get a bonus reward or not. 
In case the algorithm is correct, there is a bonus of 
$100$, from where we discount the \emph{runtime rewards} accumulated during the generation.
For example, if the agent generates a correct algorithm with a runtime reward accumulated of $-14$ during the state transitions of the generation process, the bonus will be $100+(-14)=86$. This allows the agent to receive a better bonus for the most efficient algorithms. In the case of an incorrect algorithm, the reward received by the agent will be $-1$: the number of incorrect algorithms will tend to be greater than the correct ones, so we do not want the agent to be severely penalized by finding an incorrect algorithm, since some actions of an incorrect algorithm can still lead to a correct algorithm.



\begin{table}[t]
\caption[Rewards given to each action and each condition.]{Rewards given to each action and each condition.} 
\centering 
\resizebox{9cm}{!}
{
\begin{tabular}{lr|lr}
\hline 
\textbf{Logic}  & \textbf{Reward} & \textbf{Threshold}  & \textbf{Reward} \\ 
\hline 
\action{SEND to myself}(\textless t,m\textgreater) &  $-1$ & $0$ &  $0$      \\  
\action{SEND to neighbours}(\textless t,m\textgreater) &  $-2$  & $1$ &  $-1$  \\  
\action{SEND to all}(\textless t,m\textgreater) &  $-3$ & $F+1$ &  $-2$ \\  
\action{DELIVER}(\textless m\textgreater) &  $-1$ & $(N+F)/2$ &  $-3$  \\  
\action{STOP} & $0$ & $N-F$ &  $-4$ \\  
\hline 
\end{tabular}
}
\label{table:actions_rewards} 
\label{table:conditions_rewards}
\end{table}


\subsection{Generation Process}
\label{section:generation-process}



This section explains one episode of the generation process that generates one algorithm. 

The generation process is composed of two development phases: the phase of the \event{RB-Broadcast} event handler and the phase of the \event{receive} event handler. Both phases are based on \emph{Q-Learning}\cite{watkins1992q}, one of the most adopted Reinforcement Learning algorithms. This algorithm uses a table designated \emph{QTable} to map the values of each action to each state. Furthermore, each development phase is divided in four steps: (1) {heuristic} analysis; (2) action selection; (3) reward feedback; and (4) learning update. 
Next, we explain how the development phases generate the entire algorithm during the generation process.


The generation process begins with the development phase of the RB-broadcast event handler. The agent starts with the internal state empty: \state{State([])}. Then, comes a loop-based on the current internal state, where the agent applies a set of heuristics (a topic we defer for Section \ref{section:generation_heuristics}) to discard the actions not suitable for the current state, from the set of all possible actions (step 1). 
Then, 
the agent selects one of the suitable actions based on a policy (step 2). In this work, the agent follows the \emph{Upper Confidence Bound} (UCB) policy \cite{sutton2018reinforcement}, a policy based on the idea of \emph{optimistic under uncertainty}. We selected this policy because it allows the agent to find a balance between actions less frequently chosen and actions with higher value, solving the exploration/exploitation problem\cite{coggan2004exploration}. 
Then, based on the action selected, the agent receives a runtime reward (step 3) that it uses to update its learning base, the Qtable, by associating the reward received with the action select on the current state. Moreover, the selected action is added to the current internal state and to the current event handler -- the \event{RB-Broadcast}
, originating a new state. For example, if the agent is in the state \state{State([])} and chooses action A, the action is added to the algorithm and the \event{RB-Broadcast} event handler, originating the new state \state{State([action A])}. Finally, the agent defines its current state as the new state and re-executes the development phase, returning to step 1.

The agent continues to re-execute the development phase of the \event{RB-Broadcast} event handler until the moment when it chooses the \action{STOP} action in step 2. By choosing this action, the development phase of the \event{RB-Broadcast} event handler is completed and the development phase of the \event{receive} event handler is started. The agent executes this second development phase, but now adds the actions to the \event{receive} event handler, 
until the moment when it chooses the \action{STOP} action again. This marks the end of the development phase of the \event{receive} event handler and the completion of the algorithm. The process used in the second development phase (receive) is the same as in the first (broadcast).


With the algorithm generated, the generation process ends, and the RB-Learner gives the algorithm to the RB-Oracle that, in turn, will validate it. After validating the algorithm, the RB-Oracle returns the validation result to the RB-Learner that receives a bonus, based on the algorithm being correct or not. This bonus reward will also be used to update the QTable 
of the agent.  

In the first episodes, 
the generation process will produce random algorithms, led by the policy that allows to \emph{explore} new actions and states. 
As the simulation progresses, based on the values of the QTable, 
the policy used by the agent will lead it to 
\emph{exploit} the actions that have the best value on each state, allowing it to converge to the most efficient algorithms. 
We use the terms explore and exploit with the precise meanings they have in Reinforcement Learning: explore is related with the search for new and unfamiliar states, whereas exploit refers to the examination of familiar states \cite{coggan2004exploration}. 



\subsection{Heuristics}
\label{section:generation_heuristics}



The number of possible algorithms grows exponentially with the base given by the number of actions $\mathit{T}$ (a constant), i.e., has complexity $O(\mathit{T}^i)$, where $i$ is the number of actions.
Although the agent has $\mathit{T}$ actions to choose from, there are clearly some bad choices in some cases.
For example, it is a bad option to choose 
\action{STOP} as the first action since that would generate an empty algorithm. 

To reduce the explosion of possibilities and guide the agent during the generation process, we define a set of \emph{heuristics}\cite{lenat1982nature, muller1981heuristics} for the agent to avoid bad choices. 
Notice that the heuristics do not help obtaining correct and efficient algorithms; 
they only reduce the number of possibilities to explore by discarding invalid actions in specific states and, consequently, reducing the time required to find solutions.


\begin{table}[t]
\caption[Heuristic]{Heuristics used on the generation process.} 
\centering 
\resizebox{12cm}{!}
{
\begin{tabular}{l l}
\hline 
\textbf{Heuristic}  & \textbf{Definition} \\ [0.8ex] 
\hline 
GH1 &  Do not allow repeated actions on the algorithm    \\ 
GH2 &  Allow to define the actions available in each event handler    \\ 
GH3 &  Allow to define the conditions available in each event  handler \\
GH4 &  We can only use a \action{SEND} action for each type of message sent and condition    \\ 
GH5 &  Messages sent on the \event{RB-Broadcast} event handler must be of type \type{type0}    \\ 
GH6 &  Allow to define the minimum and maximum number of  actions that each event handler can have    \\ 
GH7 &  Only select an action that waits for a message type already sent on the algorithm    \\
GH8 &  The algorithms generated must contain, at least one \action{DELIVER} action    \\
GH9 &  The incorrect states are blocked, in order to never explore them again \\
GH10 &  Allows to define the maximum number of types that the algorithm can contain \\
\hline 
\end{tabular}
}
\label{table:generation_heuristics} 
\end{table}


Table \ref{table:generation_heuristics} presents the generation heuristics (GH) that we use to guide the agent in the case of \reliablebroadcast. These heuristics were defined on the basis of a logic of discarding undesirable actions. Every time the agent is in a state, the agent uses the heuristics to know which actions are available in this specific state, thus to reduce the options from $T$ to $T_h < T$.

GH1 says that the entire algorithm cannot contain duplicate actions (except the \action{STOP} action). 
GH2 allows to define the actions available in each event handler. For \reliablebroadcast, we define that the agent can select all actions in both event handlers, except the \action{DELIVER} action on the \event{RB-Broadcast} event handler; as the \event{RB-Broadcast} event handler is only executed by one process, the \action{DELIVER} action must exist on the \event{receive} event handler, so that all processes can deliver the message, thus the possible existence of the \action{DELIVER} action on the \event{RB-Broadcast} event handler is redundant. 
GH3 allows to define the conditions available in each event handler. Based on this heuristic, we define that on the \event{receive} event handler, all considered conditions are allowed (see Section \ref{section:actions_conditions}). In the \event{RB-Broadcast}, we only allow conditions based on condition 0, since in that event handler the processes do not receive any message. 
GH4 allows to define that, for each condition and message type sent, the agent must choose between sending to all, to the neighbours or only to itself. 
GH5 allows to define a message type for the first communication step of the algorithm (\event{RB-Broadcast} event handler). 
GH6 allows to restrict the size of the algorithm generated in terms of the number of actions for each event handler. As previously explained, we took inspiration in one of the most efficient \reliablebroadcast algorithms \cite{imbs2015simple}, so we defined a minimum number of 2 and a maximum number of 4 actions in each event handler. 
GH7 forces  
to select actions based on conditions that can be validated, e.g., we forbid the agent to select actions that wait for message types not yet contained in the algorithm. 
GH8 forces the generation of 
algorithms with at least one \action{DELIVER} action, as that action clearly must exist in the algorithm.
GH9 allows to decrease the convergence time by discarding incorrect algorithms that are not related to the solution, which is equivalent to give an infinite negative reward. 
GH10 allows to define the maximum number of types that the algorithm can contain -- in this work, we defined only two possible types (cf.\ Section \ref{section:actions_message_types}).



\section{RB-Oracle}
\label{section:rb_oracle}

The RB-Oracle is the agent responsible for validating the algorithms generated by the RB-Learner agent, i.e., to implement the \emph{validation process}. This section explains the validation process executed in the context of one episode.


\LONG{\subsection{Validation Process}}

The validation process is responsible for assessing the correctness of the algorithms generated, i.e., for assessing if each algorithm satisfies the \reliablebroadcast properties (Section \ref{section:reliable_broadcast}) within one of the variants of the system model (Section \ref{section:system_model}). Every episode, the RB-Learner generates an algorithm and the RB-Oracle validates it.


Automatic validation of a fault-tolerant distributed algorithm can be achieved using different techniques such as \emph{model checking}\cite{gmeiner2014tutorial,john2013towards} or \emph{theorem proving}\cite{rahli2018velisarios}. In this paper, we use a model checking tool called
\emph{Spin}
~\cite{holzmann1997model}, a widely used framework on the validation of fault-tolerant algorithms\cite{delzanno2014,gmeiner2014tutorial,john2013towards,minamikawa2008} that allows to build models and validate them.

Spin supports a few modes. We use Spin in \emph{simulation} mode, i.e., we use it to simulate the execution of the generated algorithm in a specific system model, doing an exhaustive exploration of the state space.
In essence, during the state space exploration, Spin verifies if none of the three \reliablebroadcast properties (RB-Agreement, RB-Validity, and RB-Integrity) is violated. The three properties, the protocol, the values of $N$ and $F$, the system architecture, the behavior of each failure mode (No/Crash/Byzantine-Failure mode), the process that initiates the verification (randomly selected), and the faulty processes (also randomly selected, for $F>0$) are all specified in PROMELA (Process or Protocol Meta Language).

\LONG{\subsection{Validation of Failure Modes}}
\label{section:validation_process_failure_modes}

The RB-Oracle validates algorithms considering three failure modes: No-Failure, Crash-Failure, and Byzantine-Failure.

For the \emph{No-Failure mode}, the RB-Oracle verifies the algorithm considering that all processes are correct, i.e., by following the actions of the algorithm without deviations. In this mode, in the experiments, we assume a system with $N=3$ processes and $F=0$. Moreover, we model only one possible verification of the system: since the algorithm that runs in each process is the same, more than one model would be redundant.

In the \emph{Crash-Failure mode}, we simulate the crash of the process assuming the worst case possible: crash failures happen between the sending of messages, since the impact of a crash failure is the highest when it leads a message to be delivered only to a fraction of the processes. In this mode, in the experiments, we assume a system with $F=1$ faulty processes and $N=3$ processes, the minimum necessary to have $F=1$ failures (see Section \ref{section:failure_modes}). Moreover, for this mode we build two models: when the process that initiates the algorithm is correct and when it is faulty. This allows verifying cases when either of the event handlers fail.

For the \emph{Byzantine-Failure mode}, the RB-Oracle models a range of attacks where all faulty processes send the same malicious message to a predefined group of correct processes -- from a group with 0 processes, and consequently, not sending to anyone, to sending to $N-F$ processes, and consequently, sending to all correct processes. In this mode, in the experiments, we assume a system with $F=1$ faulty processes and $N=4$ processes, which is the minimum number of processes necessary to have $F=1$ faulty processes (see Section~\ref{section:failure_modes}). Moreover, similar to the process on the Crash-Failure mode, in this mode we also build two models: one to model a failure on each event handler of the algorithm.


\LONG{

\subsection{False-Positives}

Spin has the benefit of supporting the verification of distributed algorithms. However, model checking has the disadvantage of not allowing to prove that an algorithm is always correct, only that the properties were never violated in all states that were simulated. Moreover, it is not possible to verify all possibilities because there are infinite values of $N \in \mathbb{N}$ and many possibilities of arbitrary behavior in the Byzantine-Failure mode. Therefore, our implementation of the RB-Oracle may provide false positives, i.e., to flag as correct algorithms that are not. This means that the researcher that uses the agent to generate algorithms has to do a final round of verification once the algorithm is generated. 

\new{To increase the precision of our validation process, i.e, decrease the possible occurrence of false positives, the researcher can model new faults that break the correctness of the algorithm and add them to the validation process. However, it is important to point out that the addition of new faults on the validation model will consequently increase the total time needed to run the validation process. In this work, our experimental evaluation shows that the current approach is sufficient to generate correct \reliablebroadcast algorithms.}

}


\section{Implementation}
\label{section:implementation}

\begin{table}[t]
\caption[Number of lines of code and programming language used.]{Number of lines of code and programming language used.} 
\centering 
\resizebox{8cm}{!}
{
\begin{tabular}{l r l}
\hline 
\textbf{Module}  & \textbf{Lines of Code} & \textbf{Programming Language} \\ [0.5ex] 
\hline 
RB-Learner & $4216$ & Python3 \\
Utils & $1005$ & Python3 \\
RB-Oracle &  $573$ & Python3 and PROMELA \\ 
config & $150$ & Python3 and JSON \\
\hline 
\end{tabular}
}
\label{table:number_of_lines_of_code} 
\end{table}


As explained in earlier sections, our approach is based on two agents. 
Table~\ref{table:number_of_lines_of_code} shows the number of lines of code and the programming language used to implement the entire solution. The Utils module corresponds to the functions used by both the RB-Learner and RB-Oracle agents and the config module to the configuration files used as inputs for our system. 

The RB-Learner 
uses 
the Q-Learning algorithm 
and the UCB policy to implement the generation and learning of the algorithms. The RB-Oracle models 
a distributed system, with N processes and F faulty processes, simulating the execution of the algorithm generated using the Spin framework. 
After receiving the algorithm from the RB-Learner, the RB-Oracle creates a validation model and stores it in a PROMELA language file (\textit{.pml} extension). Then, based on the \textit{.pml} file, the RB-Oracle generates a verification file (\textit{pan.c}) -- a \textit{C} program that performs a verification of the correctness requirements for the system -- and compiles it using \textit{gcc}, generating an executable file. Lastly, the agent uses Spin to run the executable file and, with that, check the correctness of the algorithm.  

The normal functioning of our solution is, as introduced in Section \ref{section:learning_distributed_algorithms}, organized in \emph{episodes} and \emph{simulations}: one episode is composed by one generation process and one validation process, while one simulation corresponds to the entire execution with usually many episodes.


\section{Experimental Evaluation}
\label{section:experimental_evaluation}

Our experimental evaluation aims to assess the effectiveness and correctness of our solution. It will answer the following questions: 
\begin{enumerate}
    \item How many states does the agent explore until finding the first correct algorithm and converging to the most efficient algorithm?
    \item How many algorithms are generated in total, in each experiment?
    \item How many algorithms are generated until finding the first correct algorithm? 
    \item What is the proportion of correct and incorrect algorithms, from the total number of generated algorithms?
    \item How does each proposed heuristic influence the learning process?
    \LONG{\item If we change the specifications of problem, is the agent still able to generate an algorithm?}
\end{enumerate}


\begin{table}[h]
\caption[Experimental evaluation inputs.]{Experimental evaluation inputs.} 
\centering 
\resizebox{11cm}{!}
{
\begin{tabular}{ l | l }
\hline 
\textbf{Generation process inputs} & \textbf{Value} \\ [0.5ex] 
\hline 
Number of simulations &  $5$ \\
Number of episodes &  $12.000$\\
Rewards & Defined at Table \ref{table:actions_rewards}\\
Heuristics to be applied and & Configured as presented on Section \ref{section:generation_heuristics} \\
configurations associated & \\
\hline 
\textbf{Validation process inputs} & \textbf{Value} \\
\hline
Failure modes to test & No-Failure, Crash-Failure \\
& and Byzantine-Failure modes \\
Number $N$ of nodes to model & $4$\\
Fault tolerance ratios  & $F=0$ for the No-Failure, $F=\lfloor(N-1)/2\rfloor$ \\
& for the Crash-Failure and $F=\lfloor(N-1)/3\rfloor$ \\
& for the Byzantine-Failure \\
Properties to be validated & \emph{RB-Agreement}, \emph{RB-Validity} and \emph{RB-Integrity}\\
Events where faults can occur & Both \event{RB-Broadcast} and \event{receive} events\\
\hline
\end{tabular}
}
\label{table:experimental_evaluation_inputs} 
\end{table}


We ran our experiments on a single machine with 32 vCPUs, 64 GB of memory and  Debian 10. We made experiments for three cases: (1) No-Failure mode only; (2) No-Failure and Crash-Failure modes; and (3) No-Failure, Crash-Failure and Byzantine-Failure modes. Each experiment is referred to by the last failure mode of the list (the most generic): No-Failure, Crash-Failure and Byzantine-Failure experiments. All results shown in the next sections are, except when noticed, averages of $5$ simulations runs, each with $12,000$ episodes -- the minimum number of episodes that we have found to be possible for the agent to converge to the most efficient algorithms in all experiments. The $12,000$ episodes took $\pm9$ hours to run on the No-Failure experiment, $\pm3$ days to run on the Crash-Failure experiment and $\pm7$ days to run on the Byzantine-Failure experiment. This increase in time is due to the time needed for Spin to verify the models. Table \ref{table:experimental_evaluation_inputs} summarize the inputs considered for the experimental evaluation.

\subsection{States Explored}
\label{section:states_explored}

\begin{figure}
\resizebox{!}{6cm}{
\begin{tikzpicture}
\begin{axis}[
    xlabel={Episodes (in thousands)},
    ylabel={Number of states explored (in thousands)},
    legend pos=north west,
    ymajorgrids=true,
    grid style=dashed,
]
\addplot[
    color=blue,
    mark=o,
    error bars/.cd,
    y dir=both,y explicit,
    ]
    coordinates {
    (0.001,0.007)+-(0.0,0.0)
    (1,2.5300)+-(0.21787335771039102,0.21787335771039102)
    (2,2.5738)+-(0.22385566778618764,0.22385566778618764)
    (3,2.5770)+-(0.22345469339443287,0.22345469339443287)
    (4,2.6066)+-(0.2228753911942725,0.2228753911942725)
    (5,2.6114)+-(0.2238763944680189,0.2238763944680189)
    (6,2.6114)+-(0.2238763944680189,0.2238763944680189)
    (7,2.6114)+-(0.2238763944680189,0.2238763944680189)
    (8,2.6126)+-(0.22425842236134633,0.22425842236134633)
    (9,2.6134)+-(0.2249245206730471,0.2249245206730471)
    (10,2.6134)+-(0.2249245206730471,0.2249245206730471)
    (11,2.6134)+-(0.2249245206730471,0.2249245206730471)
    (12,2.6134)+-(0.2249245206730471,0.2249245206730471)
    };
    
\addplot[
    color=black,
    mark=o,
    error bars/.cd,
    y dir=both,y explicit,
    ]
    coordinates {
    (0.001,0.0070)+-(0.0,0.0)
    (1,2.9986)+-(0.0037202150475476548,0.0037202150475476548)
    (2,5.5114)+-(0.005782732917920384,0.005782732917920384)
    (3,7.6586)+-(0.007605261336732617,0.007605261336732617)
    (4,9.7062)+-(0.00897552226892675,0.00897552226892675)
    (5,11.3668)+-(0.48590550521680655,0.48590550521680655)
    (6,11.8514)+-(1.0608084841289685,1.0608084841289685)
    (7,11.9400)+-(1.2137956994486347,1.2137956994486347)
    (8,11.9400)+-(1.2137956994486347,1.2137956994486347)
    (9,11.9400)+-(1.2137956994486347,1.2137956994486347)
    (10,11.9400)+-(1.2137956994486347,1.2137956994486347)
    (11,11.9400)+-(1.2137956994486347,1.2137956994486347)
    (12,11.9400)+-(1.2137956994486347,1.2137956994486347)
    };

\addplot[
    color=red,
    mark=o,
    error bars/.cd,
    y dir=both,y explicit,
    ]
    coordinates {
    (0.001, 0.007 )+-(0.0,0.0)
    (1,2.8148 )+-(0.00762627038598, 0.00762627038598)
    (2,5.0828 )+-(0.336654125179, 0.336654125179)
    (3,7.1726 )+-(0.0054258639865, 0.0054258639865)
    (4,9.1700 )+-(0.00729383301152, 0.00729383301152)
    (5,11.1310 )+-(0.00993981891183,0.00993981891183)
    (6,13.0860 )+-(0.00871779788708,0.00871779788708)
    (7,15.0354 )+-(0.00615142259969, 0.00615142259969)
    (8,16.9652 )+-(0.00222710574513,0.00222710574513)
    (9,18.9064 )+-(0.00174355957742,0.00174355957742)
    (10,19.6098 )+-(0.121061967603,0.121061967603)
    (11,19.6098 )+-(0.121061967603,0.121061967603)
    (12,19.6098 )+-(0.121061967603,0.121061967603)
    };
    
    \legend{No-Failure, Crash-Failure, Byzantine-Failure}
    
\end{axis}
\end{tikzpicture}
}
\centering
\caption{Number of states explored during each experiment.}
\label{fig:number_of_states_explored}
\end{figure}
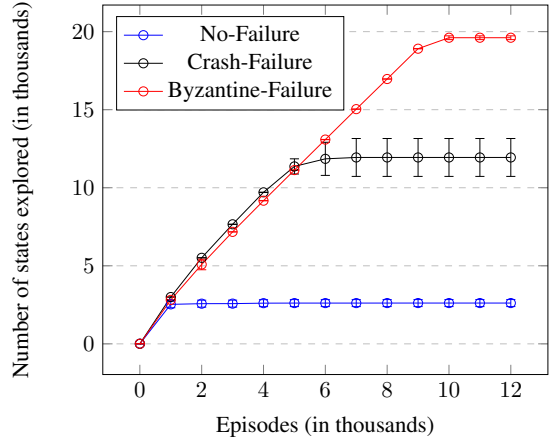

For each algorithm generated, the agent explores multiple states when selecting the actions. This first set of experiments assesses the number of states explored in each experiment. Figure \ref{fig:number_of_states_explored} shows the total number of states explored by the agent for each episode, on the entire experiment. Table \ref{table:states_explored_unit_first_valid_algorithm} shows the number of states explored until the agent generated the first correct algorithm. 

As expected, we can see in both figures that the agent needs to explore more states when the complexity of the problem to solve increases, i.e., the agent needs to explore more states when trying to find a Byzantine-tolerant algorithm -- almost $20,000$ states -- when compared to a Crash-tolerant or a non-fault-tolerant algorithm -- around $12,000$ and $3,000$ states, respectively. Additionally, the agent also takes more time to converge when trying to find a Byzantine-tolerant algorithm -- between $8,000$ and $10,000$ episodes -- when compared to the other cases -- between $1,000$ and $2,000$ episodes for the No-Failure algorithm and between $4,000$ and $6,000$ episodes for the Crash-Failure algorithm.



\begin{table}[t]
\caption[Number of states generated until the first correct algorithm is generated.]{Number of states generated until the first correct algorithm is generated.} 
\centering 
\resizebox{5.5cm}{!}
{
\begin{tabular}{lc}
\hline 
\textbf{Experiment}  & \textbf{States} \\ [0.5ex] 
\hline 
No-Failure & $15.4\pm0.0$  \\ 
Crash-Failure &  $33.4\pm16.9$   \\ 
Byzantine-Failure &  $13462.8\pm6909.4$  \\ 
\hline 
\end{tabular}
}
\label{table:states_explored_unit_first_valid_algorithm} 
\end{table}


\subsection{Algorithms Generated}
\label{section:algorithms_generated}


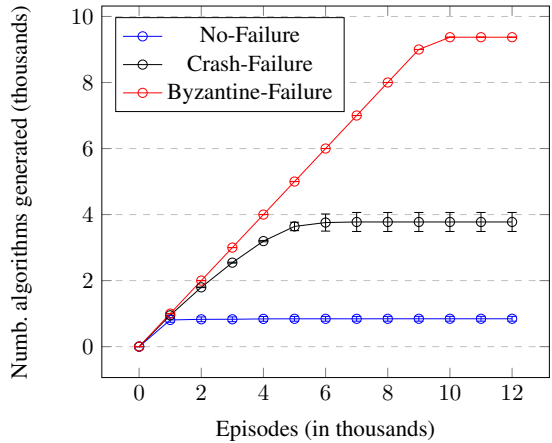
\begin{figure}[h]
\resizebox{!}{6cm}{
\begin{tikzpicture}
\begin{axis}[
    xlabel={Episodes (in thousands) },
    ylabel={Numb.\ algorithms generated (thousands)},
    legend pos=north west,
    ymajorgrids=true,
    grid style=dashed,
]
\addplot[
    color=blue,
    mark=o,
    error bars/.cd,
    y dir=both,y explicit,
    ]
    coordinates {
    (0.001,0.001)+-(0.0,0.0)
    (1,0.8076)+-(0.06960632155199698,0.06960632155199698)
    (2,0.8276)+-(0.07202110801702511,0.07202110801702511)
    (3,0.8286)+-(0.07202110801702511,0.07202110801702511)
    (4,0.8404)+-(0.07189603605206618,0.07189603605206618)
    (5,0.8428)+-(0.07241104888067842,0.07241104888067842)
    (6,0.8428)+-(0.07241104888067842,0.07241104888067842)
    (7,0.8428)+-(0.07241104888067842,0.07241104888067842)
    (8,0.8434)+-(0.0725936636353339,0.0725936636353339)
    (9,0.8438)+-(0.07294765246394157,0.07294765246394157)
    (10,0.8438)+-(0.07294765246394157,0.07294765246394157)
    (11,0.8438)+-(0.07294765246394157,0.07294765246394157)
    (12,0.8438)+-(0.07294765246394157,0.07294765246394157)
    };
\addplot[
    color=black,
    mark=o,
    error bars/.cd,
    y dir=both,y explicit,
    ]
    coordinates {
    (0.001,0.001)+-(0.0,0.0)
    (1,0.9488)+-(0.0031874754901018454,0.0031874754901018454)
    (2,1.7960)+-(0.008318653737234168,0.008318653737234168)
    (3,2.5432)+-(0.013526270735128733,0.013526270735128733)
    (4,3.1960)+-(0.021080796948882174,0.021080796948882174)
    (5,3.6406)+-(0.12835201595611966,0.12835201595611966)
    (6,3.7582)+-(0.25893582216448925,0.25893582216448925)
    (7,3.7758)+-(0.28784120622315356,0.28784120622315356)
    (8,3.7758)+-(0.28784120622315356,0.28784120622315356)
    (9,3.7758)+-(0.28784120622315356,0.28784120622315356)
    (10,3.7758)+-(0.28784120622315356,0.28784120622315356)
    (11,3.7758)+-(0.28784120622315356,0.28784120622315356)
    (12,3.7758)+-(0.28784120622315356,0.28784120622315356)
    };
    
\addplot[
    color=red,
    mark=o,
    error bars/.cd,
    y dir=both,y explicit,
    ]
    coordinates {
    (0.001,0.001)+-(0.0,0.0)
    (1,1)+-(0.0,0.0)
    (2,2)+-(0.0,0.0)
    (3,3)+-(0.0,0.0)
    (4,4)+-(0.0,0.0)
    (5,5)+-(0.0,0.0)
    (6,6)+-(0.0,0.0)
    (7,7)+-(0.0,0.0)
    (8,8)+-(0.0,0.0)
    (9,9)+-(0.0,0.0)
    (10,9.3716)+-(0.0068,0.0068)
    (11,9.3716)+-(0.0068,0.0068)
    (12,9.3716)+-(0.0068,0.0068)
    };
    
    \legend{No-Failure, Crash-Failure, Byzantine-Failure}
    
\end{axis}
\end{tikzpicture}
}
\centering
\caption{Number of algorithms generated during each experiment.}
\label{fig:number_of_algorithms_generated}
\end{figure}


\begin{table}[h]
\caption[Number of correct and incorrect algorithms generated in each experiment and number of algorithms generated until the first correct algorithm.]{Number of correct and incorrect algorithms generated in each experiment and number of algorithms generated until the first correct algorithm.} 
\centering 
\resizebox{10cm}{!}
{
\begin{tabular}{lcc|c}
\hline 
\textbf{Experiment}  & \textbf{Correct Algo.}& \textbf{Incorrect Algo.} & \textbf{Number Algo.} \\ [0.5ex] 
\hline 
No-Fail. &  $341.8\pm23.2$ & $502.0\pm53.3$ & $2.6\pm1.9$ \\ 
Crash-Fail. &  $420.2\pm32.1$ & $3355.6\pm256.6$ & $6.8\pm4.1$  \\ 
Byzantine-Fail. &  $2.0\pm0.0$ & $9369.6\pm6.8$ & $6260.0\pm3467.9$\\ 
\hline 
\end{tabular}
}
\label{table:valid_and_invalid_algortihms_generated} 
\label{table:number_of_algorithm_generated_until_first_valid}
\end{table}


The agent generates multiple algorithms with the objective to learn from them and, therefore, we decided to assess how many algorithms the agent generates in each experiment. Figure \ref{fig:number_of_algorithms_generated} shows the number of algorithms generated by the agent per episode. Table \ref{table:number_of_algorithm_generated_until_first_valid} shows the number of correct and incorrect algorithms generated, as also the number of algorithms generated until the first correct algorithm. 
As expected, and similarly to what happens with the number of states, the agent needs to generate more algorithms, as also takes more time to converge, with the increase of the complexity of the problem to solve. Moreover, another interesting aspect is the percentage of incorrect algorithms generated by the agent in each test: $\pm60$\% on the No-Failure test, $\pm89$\% on the Crash-Failure test and $99.9$\% on the Byzantine-Failure test, which means that, even with all the Heuristics defined, the agent still has a difficult task to generate a correct algorithm.

The final algorithms generated by each experiment are presented in Algorithms \ref{alg:best_NoFailure_RB_algorithm_generated}, \ref{alg:best_CrashFailure_RB_algorithm_generated} and \ref{alg:best_ByzantineFailure_RB_algorithm_generated}, 
one for each failure mode. 


\begin{algorithm}[ht]
\caption{Most efficient \reliablebroadcast algorithm for a No-Failure experiment generated by the RB-Learner.}\label{alg:best_NoFailure_RB_algorithm_generated}
\begin{algorithmic}[1]\small
\BState \emph{when RB-Broadcast(m) do}:
\BState \hskip1.0em \action{SEND to all}(\textless \type{type0},m\textgreater) if received (\textless \type{type0},m\textgreater) from $0$ distinct parties and not already sent;
\State \hskip1.0em \action{STOP} if received (\textless \type{type0},m\textgreater) from $0$ distinct parties;
\newline
\BState \emph{when receive(m) do}:
\State \hskip1.0em \action{DELIVER}(\textless m\textgreater) if received (\textless \type{type0},m\textgreater) from $0$ distinct parties and not already delivered;
\State \hskip1.0em \action{STOP} if received (\textless \type{type0},m\textgreater) from $0$ distinct parties;
\end{algorithmic}
\end{algorithm}


In the No-Failure mode, the agent converged to Algorithm \ref{alg:best_NoFailure_RB_algorithm_generated} in 4 of the simulations executed. 
This algorithm is equivalent to one presented in \cite{cachin2011introduction}: both exchange, at most, $N$ messages, require $1$ communication step and $1$ message type (or none) and need to receive $1$ message.


\begin{algorithm}[t]
\caption{Most efficient \reliablebroadcast algorithm for a Crash-Failure experiment generated by the RB-Learner.}\label{alg:best_CrashFailure_RB_algorithm_generated}
\begin{algorithmic}[1]\small
\BState \emph{when RB-Broadcast(m) do}:
\BState \hskip1.0em \action{SEND to myself}(\textless \type{type0},m\textgreater) if received (\textless \type{type0},m\textgreater) from $0$ distinct parties and not already sent;
\State \hskip1.0em \action{STOP} if received (\textless \type{type0},m\textgreater) from $0$ distinct parties;
\newline
\BState \emph{when receive(m) do}:
\State \hskip1.0em \action{SEND to neighbours}(\textless \type{type1},m\textgreater) if received (\textless \type{type0},m\textgreater) from $0$ distinct parties and not already sent;
\State \hskip1.0em \action{DELIVER}(\textless m\textgreater) if received (\textless \type{type0},m\textgreater) from $0$ distinct parties and not already delivered;
\State \hskip1.0em \action{STOP} if received (\textless \type{type0},m\textgreater) from $0$ distinct parties;
\end{algorithmic}
\end{algorithm}


On the Crash-Failure mode, the agent also converged to Algorithm \ref{alg:best_CrashFailure_RB_algorithm_generated} in 4 of the simulations executed. 
Note that the algorithm sends a new message type on the \event{receive} event handler (\type{type1}), when it could send the \type{type0}. This happens because of the heuristic GH5, that only allows to send messages of \type{type0} on the \event{RB-Broadcast} event handler. This algorithm is similar to the one presented in \cite{hadzilacos1994modular,raynal2018fault}: both exchange, at most, $N^2-N+1$ messages, require $1$ communication step and $1$ message type (or none) and need to receive $1$ message.

On the Byzantine-Failure mode, the agent converged to Algorithm \ref{alg:best_ByzantineFailure_RB_algorithm_generated} in all the simulations executed. 
This algorithm is one of the most efficient algorithms developed and it is similar to the one presented in \cite{imbs2015simple}: both exchange, at most, $N^2+N$ messages, require $2$ communication steps and $2$ message types and need to receive $(N+F)/2$ messages.

\begin{algorithm}
\caption{Most efficient \reliablebroadcast algorithm for a Byzantine-Failure experiment generated by the RB-Learner.}\label{alg:best_ByzantineFailure_RB_algorithm_generated}
\begin{algorithmic}[1]\small
\BState \emph{when RB-Broadcast(m) do}:
\State \hskip1.0em \action{SEND to all}(\textless \type{type0},m\textgreater) if received (\textless \type{type0},m\textgreater) from $0$ distinct parties and not already sent;
\State \hskip1.0em \action{STOP} if received (\textless \type{type0},m\textgreater) from $0$ distinct parties;
\newline
\BState \emph{when receive(m) do}:
\State \hskip1.0em \action{SEND to all}(\textless \type{type1},m\textgreater) if received (\textless \type{type0},m\textgreater) from $1$ distinct party and not already sent;
\State \hskip1.0em \action{DELIVER}(\textless m\textgreater) if received (\textless \type{type1},m\textgreater) from $(N+F)/2$ distinct parties and not already delivered;
\State \hskip1.0em \action{SEND to all}(\textless \type{type1},m\textgreater) if received (\textless \type{type1},m\textgreater) from $F+1$ distinct parties and not already sent;
\State \hskip1.0em \action{STOP} if received (\textless \type{type0},m\textgreater) from $0$ distinct parties;
\end{algorithmic}
\end{algorithm}

\subsection{Impact of the Heuristics}

The heuristics we defined (see Section \ref{section:generation_heuristics}) guide the agent by helping it to avoid incorrect algorithms
\SHORT{ and reduce the number of states to explore.}
\LONG{.They do not help obtaining algorithms or algorithms that are more correct, but reduce the number of states to explore.}
In this evaluation, we analyzed the importance of each heuristic with the Crash-Failure experiment.

To achieve this, we ran one experiment with each GH turned off and all others turned on. There were two exceptions. In GH6, we  increased the maximum number of actions in each event from $4$ to $5$, but did not turn this heuristic off, to avoid that the agent would generate algorithms with too many actions. For GH10, we increased the maximum number of types from $2$ to $3$ but did not turn it off, as the agent could explore too many types. 
We executed one simulation with $10,000$ episodes for each experiment. 

\begin{figure}
\resizebox{!}{6cm}{
\begin{tikzpicture}
\begin{axis}[
    xlabel={Episodes (in thousands) },
    ylabel={Num.\ algorithms generated (thousands)},
    legend pos= outer north east,
    ymajorgrids=true,
    grid style=dashed,
]
\addplot[
    color=black,
    mark=o,
    error bars/.cd,
    y dir=both,y explicit,
    ]
    coordinates {
    (0.001,0.001)+-(0.0,0.0)
    (1,0.9488)+-(0.0031874754901018454,0.0031874754901018454)
    (2,1.7960)+-(0.008318653737234168,0.008318653737234168)
    (3,2.5432)+-(0.013526270735128733,0.013526270735128733)
    (4,3.1960)+-(0.021080796948882174,0.021080796948882174)
    (5,3.6406)+-(0.12835201595611966,0.12835201595611966)
    (6,3.7582)+-(0.25893582216448925,0.25893582216448925)
    (7,3.7758)+-(0.28784120622315356,0.28784120622315356)
    (8,3.7758)+-(0.28784120622315356,0.28784120622315356)
    (9,3.7758)+-(0.28784120622315356,0.28784120622315356)
    (10,3.7758)+-(0.28784120622315356,0.28784120622315356)
    };
    
\addplot[
    color=gray,
    mark=x,
    error bars/.cd,
    y dir=both,y explicit,
    ]
    coordinates {
    (0.0001,0.0001)+-(0.0,0.0)
    (1,0.9980)+-(0.0,0.0)
    (2,1.9860)+-(0.0,0.0)
    (3,2.9640)+-(0.0,0.0)
    (4,3.8820)+-(0.0,0.0)
    (5,4.1720)+-(0.0,0.0)
    (6,4.1760)+-(0.0,0.0)
    (7,4.1760)+-(0.0,0.0)
    (8,4.1760)+-(0.0,0.0)
    (9,4.1760)+-(0.0,0.0)
    (10,4.1760)+-(0.0,0.0)
    };

\addplot[
    color=red,
    mark=triangle,
    error bars/.cd,
    y dir=both,y explicit,
    ]
    coordinates {    
    (0.0001,0.0001)+-(0.0,0.0)
    (1,1)+-(0.0,0.0)
    (2,1.9990)+-(0.0,0.0)
    (3,2.9970)+-(0.0,0.0)
    (4,3.9900)+-(0.0,0.0)
    (5,4.9810)+-(0.0,0.0)
    (6,5.9740)+-(0.0,0.0)
    (7,6.9470)+-(0.0,0.0)
    (8,7.9240)+-(0.0,0.0)
    (9,8.8820)+-(0.0,0.0)
    (10,9.4370)+-(0.0,0.0)
    };

\addplot[
    color=blue,
    mark=diamond,
    error bars/.cd,
    y dir=both,y explicit,
    ]
    coordinates {    
    (0.0001,0.0001)+-(0.0,0.0)
    (1,1)+-(0.0,0.0)
    (2,2)+-(0.0,0.0)
    (3,2.9990)+-(0.0,0.0)
    (4,3.9990)+-(0.0,0.0)
    (5,4.9970)+-(0.0,0.0)
    (6,5.9970)+-(0.0,0.0)
    (7,6.9960)+-(0.0,0.0)
    (8,7.996)+-(0.0,0.0)
    (9,8.994)+-(0.0,0.0)
    (10,9.9920)+-(0.0,0.0)
    };

\addplot[
    color=green,
    mark=pentagon,
    error bars/.cd,
    y dir=both,y explicit,
    ]
    coordinates {    
    (0.0001,0.00010)+-(0.0,0.0)
    (1,0.9990)+-(0.0,0.0)
    (2,1.9940)+-(0.0,0.0)
    (3,2.9780)+-(0.0,0.0)
    (4,3.9550)+-(0.0,0.0)
    (5,4.9070)+-(0.0,0.0)
    (6,5.6140)+-(0.0,0.0)
    (7,5.6460)+-(0.0,0.0)
    (8,5.6540)+-(0.0,0.0)
    (9,5.6540)+-(0.0,0.0)
    (10,5.6540)+-(0.0,0.0)
    };
    
\addplot[
    color=yellow,
    mark=asterisk,
    error bars/.cd,
    y dir=both,y explicit,
    ]
    coordinates {    
    (0.0001,0.0001)+-(0.0,0.0)
    (1,1.0)+-(0.0,0.0)
    (2,1.9990)+-(0.0,0.0)
    (3,2.9960)+-(0.0,0.0)
    (4,3.9930)+-(0.0,0.0)
    (5,4.9860)+-(0.0,0.0)
    (6,5.9700)+-(0.0,0.0)
    (7,6.9490)+-(0.0,0.0)
    (8,7.9200)+-(0.0,0.0)
    (9,8.8750)+-(0.0,0.0)
    (10,9.8080)+-(0.0,0.0)
    };

\addplot[
    color=lime,
    mark=square,
    error bars/.cd,
    y dir=both,y explicit,
    ]
    coordinates {    
    (0.0001,0.00010)+-(0.0,0.0)
    (1,1.0)+-(0.0,0.0)
    (2,1.9970)+-(0.0,0.0)
    (3,2.9920)+-(0.0,0.0)
    (4,3.9860)+-(0.0,0.0)
    (5,4.3200)+-(0.0,0.0)
    (6,4.3450)+-(0.0,0.0)
    (7,4.3470)+-(0.0,0.0)
    (8,4.3490)+-(0.0,0.0)
    (9,4.3570)+-(0.0,0.0)
    (10,4.3690)+-(0.0,0.0)
    };
   
\addplot[
    color=orange,
    mark=*,
    error bars/.cd,
    y dir=both,y explicit,
    ]
    coordinates {    
    (0.0001,0.0001)+-(0.0,0.0)
    (1,0.9990)+-(0.0,0.0)
    (2,1.9880)+-(0.0,0.0)
    (3,2.9720)+-(0.0,0.0)
    (4,3.9510)+-(0.0,0.0)
    (5,4.9060)+-(0.0,0.0)
    (6,5.8020)+-(0.0,0.0)
    (7,6.5940)+-(0.0,0.0)
    (8,6.7730)+-(0.0,0.0)
    (9,6.7730)+-(0.0,0.0)
    (10,6.7730)+-(0.0,0.0)
    };
   
\addplot[
    color=pink,
    mark=otimes,
    error bars/.cd,
    y dir=both,y explicit,
    ]
    coordinates {
    (0.0001,0.0001)+-(0.0,0.0)
    (1,0.9970)+-(0.0,0.0)
    (2,1.9900)+-(0.0,0.0)
    (3,2.9710)+-(0.0,0.0)
    (4,3.9220)+-(0.0,0.0)
    (5,4.3740)+-(0.0,0.0)
    (6,4.4000)+-(0.0,0.0)
    (7,4.4060)+-(0.0,0.0)
    (8,4.4100)+-(0.0,0.0)
    (9,4.4130)+-(0.0,0.0)
    (10,4.4140)+-(0.0,0.0)
    }; 

\addplot[
    color=brown,
    mark=heart,
    error bars/.cd,
    y dir=both,y explicit,
    ]
    coordinates {
    (0.0001,0.00010)+-(0.0,0.0)
    (1,0.9630)+-(0.0,0.0)
    (2,1.8480)+-(0.0,0.0)
    (3,2.6330)+-(0.0,0.0)
    (4,3.3070)+-(0.0,0.0)
    (5,3.8770)+-(0.0,0.0)
    (6,4.3360)+-(0.0,0.0)
    (7,4.3360)+-(0.0,0.0)
    (8,4.3360)+-(0.0,0.0)
    (9,4.3360)+-(0.0,0.0)
    (10,4.3360)+-(0.0,0.0)
    }; 
 
\addplot[
    color=purple,
    mark=Mercedes star,
    error bars/.cd,
    y dir=both,y explicit,
    ]
    coordinates {
    (0.0001,0.0001)+-(0.0,0.0)
    (1,1.0)+-(0.0,0.0)
    (2,1.997)+-(0.0,0.0)
    (3,2.9960)+-(0.0,0.0)
    (4,3.9880)+-(0.0,0.0)
    (5,4.978)+-(0.0,0.0)
    (6,5.968)+-(0.0,0.0)
    (7,6.9520)+-(0.0,0.0)
    (8,7.9130)+-(0.0,0.0)
    (9,8.8350)+-(0.0,0.0)
    (10,9.2190)+-(0.0,0.0)
    }; 
    
\legend{All,GH1,GH2,GH3,GH4,GH5,GH6,GH7,GH8,GH9,GH10}
\end{axis}
\end{tikzpicture}
}
\centering
\caption{Number of algorithms generated during each experiment without the identified GH. The All line represents a experiment with all GH turned on.}
\label{fig:number_of_algorithms_generated_testing_heuristics}
\end{figure}
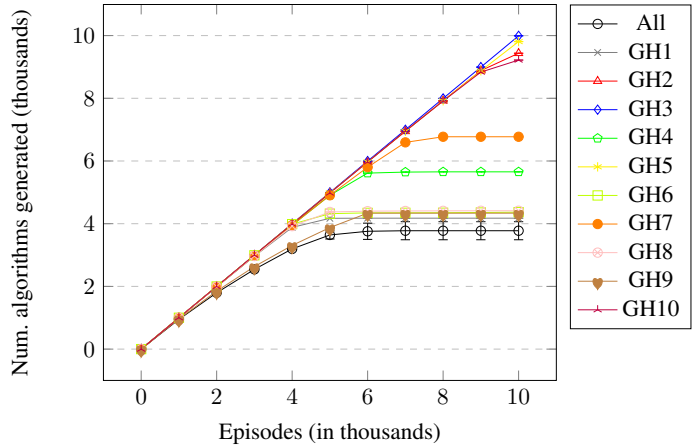

Figure~\ref{fig:number_of_algorithms_generated_testing_heuristics} shows the evolution of the number of algorithms generated with each GH turned off, 
\LONG{and Table \ref{table:valid_and_invalid_algortihms_generated_testing_heuristics} shows the number of correct and incorrect algorithms generated and the number of algorithms generated until the first correct algorithm.}
%
from where we conclude that all heuristics are important to reduce the number of states explored until a correct and efficient algorithm is obtained.



\LONG{
\begin{table}[t]
\caption[Number of correct and incorrect algorithms generated in each experiment and number of algorithms generated until the first correct algorithm.]{Number of correct and incorrect algorithms generated in each experiment and number of algorithms generated until the first correct algorithm.} 
\centering 
\begin{tabular}{l r r |r}
\hline 
\textbf{Experiment}  & \textbf{Correct Algo.}& \textbf{Incorrect Algo.} & \textbf{Number Algo.}\\ [0.5ex] 
\hline 
all GH &  $420\pm32.1$ & $3356\pm256.6$ & $6.8\pm4.1$ \\
w/o GH1 &  $426\pm0.0$ & $3750\pm0.0$ &  $8.0\pm0.0$ \\
w/o GH2 &  $523\pm0.0$ & $8914\pm0.0$ & $6.0\pm0.0$\\
w/o GH3 &  $684\pm0.0$ & $9308\pm0.0$ & $14.0\pm0.0$ \\
w/o GH4 &  $673\pm0.0$ & $4981\pm0.0$ & $3.0\pm0.0$  \\
w/o GH5 &  $1660\pm0.0$ & $8148\pm0.0$ &  $8.0\pm0.0$ \\
w/o GH6 &  $917\pm0.0$ & $3452\pm0.0$ & $12.0\pm0.0$  \\
w/o GH7 &  $488\pm0.0$ & $6285\pm0.0$ & $7.0\pm0.0$ \\
w/o GH8 &  $272\pm0.0$ & $4142\pm0.0$ & $8.0\pm0.0$  \\
w/o GH9 &  $458\pm0.0$ & $3878\pm0.0$ & $5.0\pm0.0$ \\
w/o GH10 &  $878\pm0.0$ & $8341\pm0.0$ & $6.0\pm0.0$ \\
\hline 
\end{tabular}
\label{table:valid_and_invalid_algortihms_generated_testing_heuristics} 
\end{table}
}





\LONG{

\subsection{Modifying the Problem}

This section evaluates the last question, i.e., if the agent is still able to generate an algorithm if the properties of the problem are modified. The change we make is as follows.

\new{The optimal tolerance ratio of the Byzantine-Failure mode is $\lfloor(N-1)/3\rfloor$. Since in this mode crash failures can occur -- the Byzantine-Failure mode considers arbitrary behavior on the faulty processes -- we have decided to change the properties of the Crash-Failure mode by decreasing the faults tolerated from $\lfloor(N-1)/2\rfloor$ to $\lfloor(N-1)/3\rfloor$, but maintaining the same Byzantine tolerance ratio.}


Algorithm~\ref{alg:new_crash_tolerant_algorithm} shows \new{the new algorithm generated by the RB-Learner that is $\lfloor(N-1)/3\rfloor$ Byzantine-tolerant and $\lfloor(N-1)/3\rfloor$ Crash-tolerant.} 
When compared with algorithm \ref{alg:best_ByzantineFailure_RB_algorithm_generated} -- both were tested with the same Byzantine-tolerance ratio but a different Crash-tolerance ratio -- the new algorithm presents two improvements: the algorithm sends less messages and needs to wait for less messages.

The first improvement is minimal: it comes from the algorithm sending messages only to the neighbors, instead of sending to all. With this, each process spares the cost of sending a message to itself, allowing to save a total $N+1$ messages, i.e., $1$ message of \type{type0} and $N$ messages of \type{type1}. 

The second improvement is also influenced by the first one, since, by sending only to the neighbours, this allows the algorithm to wait for only $F+1$ instead of $(N+F)/2$ messages to deliver the message. This new threshold 
can have a considerable impact when waiting for messages from other processes. For example, if we had a system with $N=100$ and $F=33$, waiting for 
$(N+F)/2$ messages 
costs 
$(100+33)/2=66.5\approx67$ messages, while 
waiting for $F+1$ messages 
costs only $33+1=34$ messages, which presents a reduction of almost $50\%$ on the number of messages needed.

This experiment demonstrates the capability of the RB-Learner to adapt to the specified problem with modifications and learn to generate the most efficient algorithm for that case also. 


\begin{algorithm}
\caption{Byzantine-tolerant algorithm generated by the RB-Leaner with the new crash-tolerance assumption.}\label{alg:new_crash_tolerant_algorithm}
\begin{algorithmic}[1]\small
\BState \emph{when RB-Broadcast(m) do}:
\State \hskip1.0em \action{SEND to neighbours}(\textless \type{type0},m\textgreater) if received (\textless \type{type0},m\textgreater) from $0$ distinct parties and not already sent;
\State \hskip1.0em \action{STOP} if received (\textless \type{type0},m\textgreater) from $0$ distinct parties;
\newline
\BState \emph{when receive(m) do}:
\State \hskip1.0em \action{SEND to neighbours}(\textless \type{type1},m\textgreater) if received (\textless \type{type0},m\textgreater) from $1$ distinct parties and not already sent;
\State \hskip1.0em \action{SEND to neighbours}(\textless \type{type1},m\textgreater) if received (\textless \type{type1},m\textgreater) from $F+1$ distinct parties and not already sent;
\State \hskip1.0em \action{DELIVER}(\textless m\textgreater) if received (\textless \type{type1},m\textgreater) from $F+1$ distinct parties and not already delivered;
\State \hskip1.0em \action{STOP} if received (\textless \type{type0},m\textgreater) from $0$ distinct parties;
\end{algorithmic}
\end{algorithm}

}


\section{Related Work}
\label{section:related_work}

Fault-tolerant algorithms have been widely studied over the years\cite{anema2021message,ben1983another,bracha1984asynchronous,bonomi2021practical,castro1999practical,chang1984reliable,correia2006consensus,correia2010asynchronous,imbs2015simple,imbs2016trading,raynal2018fault}.
These algorithms: 
solve different problems, such as \emph{Reliable Broadcast} 
\cite{imbs2015simple} and \emph{Consensus}\cite{castro1999practical}; 
tolerate different failure modes, like \emph{Crash}\cite{hadzilacos1994modular} and \emph{Byzantine}\cite{correia2010asynchronous}; 
use different communication models, namely \emph{fully-connected}\cite{correia2006consensus} and \emph{partially-connected}\cite{bonomi2021practical};
and tolerate different fault ratios, such as $\lfloor(N-1)/3\rfloor$\cite{bracha1984asynchronous} and $\lfloor(N-1)/2\rfloor$\cite{correia2010asynchronous}, where N is the number of components in the system. 
However, as far as we know, all works are based on manual or brute-force processes, without any kind of 
artificial intelligence helping with the process of generating the algorithms.

Automatic code generation works focused on local, non-distributed, single-threaded code, started by using static techniques such as design-based approaches\cite{budinsky1996automatic}, UML\cite{moreira2010automatic} or reverse engineering\cite{orvalho2020squares},
but lately has been focused on using machine learning techniques\cite{allamanis2018survey,acsirouglu2019automatic,zhang2019analysis} 
mostly 
supervised 
learning techniques, such as Deep Learning\cite{le2020deep,zhu2021code}. For distributed code, we identified two works: one that automatically finds mutual exclusion algorithms\cite{bar2003automatic} and another that automatically investigates and validates Consensus algorithms\cite{zielinski2007automatic}, both using brute-force approaches. However, in our case, we use an unsupervised machine learning technique -- Reinforcement Learning~\cite{kaelbling1996reinforcement,sutton2018reinforcement,wiering2012reinforcement}  -- that allows the agent to learn 
without the need of prior knowledge about solutions to the problem to be solved. Reinforcement Learning has been explored mainly in games \cite{lample2017playing,mnih2013playing,silver2018general} and robotics\cite{ibarz2021train,kober2013reinforcement,mataric1997reinforcement},  so our work applies it to an entirely different problem.
When compared to our approach, the most interesting works that use this technique are those that use an agent to generate GPU compiler heuristics\cite{colbert2021generating}, an agent that chooses the most suitable algorithm\cite{lagoudakis2000algorithm}, or an agent that is capable of generating experiment input data\cite{kim2018generating}. Nevertheless, these problems are very different from ours.

For the validation of fault-tolerant algorithms 
\cite{gmeiner2014tutorial,goel2021towards,lamport1994specifying,rahli2018velisarios}, we have identified and used the Spin/PROMELA~\cite{holzmann1997model} model checker\cite{gmeiner2014tutorial,john2013towards} framework, since it allows us to model and validate the generated algorithms, is used by a significant number of related works\cite{delzanno2014,gmeiner2014tutorial,john2013towards,minamikawa2008} and also has good community support resources. Besides Spin/PROMELA, other possible languages and frameworks could have been used: TLA+\footnote{\url{https://lamport.azurewebsites.net/tla/tla.html}} \cite{lamport1994specifying}, 
the ByMC\footnote{\url{https://github.com/konnov/bymc}} framework\cite{konnov2018bymc} and the IC3PO\footnote{\url{https://github.com/aman-goel/ic3po}}\cite{goel2021ic3po}.


\section{Conclusion}
\label{section:conclusion}

Fault-tolerant algorithms have been studied over the years, discussing different problems and variants. However, this study is complex and was always based on a human-oriented process.

To automate this process, we propose a solution based on two agents,  RB-Learner and  RB-Oracle, capable of learning to generate a distributed algorithm, more precisely the \reliablebroadcast algorithm
. As we have presented during the experimental evaluation, our solution is capable of generating correct and efficient algorithms, depending on the properties of the problem, which proves that our approach can be used to generate distributed algorithms. 

To the best of our knowledge, this work is the first that merges both areas of generation and validation into an automatic process capable of generating correct and efficient \reliablebroadcast algorithms. 
Additionally, this is the first work that uses a machine learning approach to 
generate correct and efficient algorithms to solve a specific distributed problem.

For further research, we aim to apply our approach to different distributed problems, e.g. Consensus, and try to decrease the number of inputs needed, to further decouple our agent from knowledge based on previous works, e.g. the threshold expressions.



\bibliographystyle{plainurl}
\bibliography{paper}  






\end{document}